\documentclass{article}
\usepackage{arxiv}

\usepackage{xcolor}  

\usepackage{amssymb}

\usepackage{amsmath}
\usepackage{mathdots}

\usepackage{mathtools} 

\usepackage{tensor}

\usepackage{urwchancal}
\DeclareFontFamily{OT1}{pzc}{}
\DeclareFontShape{OT1}{pzc}{m}{it}{<-> s * [1.10] pzcmi7t}{}
\DeclareMathAlphabet{\mathpzc}{OT1}{pzc}{m}{it}

\usepackage{graphicx}
\usepackage{ifpdf}
\ifpdf
\usepackage{epstopdf}
\PrependGraphicsExtensions{.svg}
\fi

\usepackage{amsthm}
\newtheorem{theorem}{Theorem}[section]

\newtheorem{corollary}[theorem]{Corollary}
\newtheorem{proposition}[theorem]{Proposition}

\providecommand{\R}{\mathbb{R}}

\providecommand{\SO}{\mathbf{SO}}

\providecommand{\SE}{\mathbf{SE}}

\providecommand{\grpG}{\mathbf{G}}

\providecommand{\gothg}{\mathfrak{g}}

\providecommand{\gothX}{\mathfrak{X}} 

\providecommand{\se}{\mathfrak{se}}

\providecommand{\calM}{\mathcal{M}}
\providecommand{\calN}{\mathcal{N}}

\providecommand{\calU}{\mathcal{U}}

\providecommand{\vecL}{\mathbb{L}}

\providecommand{\Sym}{\mathbb{S}} 

\DeclareMathOperator{\Ad}{Ad}

\providecommand{\id}{\mathrm{id}} 

\providecommand{\tT}{\mathrm{T}} 
\providecommand{\td}{\mathrm{d}}
\providecommand{\tD}{\mathrm{D}}

\providecommand{\ddt}{\frac{\td}{\td t}}

\providecommand{\mr}[1]{{#1}^\circ} 

\providecommand{\ob}[1]{\overline{#1}} 
\usepackage{accents}
\makeatletter
\providecommand{\scirc}{%
    \hbox{\fontfamily{\rmdefault}\fontsize{0.4\dimexpr(\f@size pt)}{0}\selectfont{\raisebox{-0.52ex}[0ex][-0.52ex]{$\circ$}}}}

\makeatother

\mathchardef\mhyphen="2D

\usepackage{hyperref}

\usepackage{tikz}
\usetikzlibrary{positioning}

\usepackage{graphicx}      
\usepackage{natbib} 

\usepackage{xcolor}

\usepackage[all]{xy}

\newcommand\at[2]{\left.#1\right|_{#2}}
\newcommand{\dds}{\frac{\td}{\td s}}
\newcommand{\figsize}{0.7}

\begin{document}

\newcommand{\papercitation}{
Under Review.
}
\newcommand{\linktopaper}{
https://arxiv.org/}
\newcommand{\DOInumber}{
}
\newcommand{\publicationdetails}
{\copyrightNoticeIFACAccepted{2022}
This research was supported by the Australian Research Council through the Discovery Project DP210102607.
}
\newcommand{\publicationversion}
{Preprint}

\title{Equivariant Filters are Equivariant}
\headertitle{Equivariant Filters are Equivariant}

\author{
\href{https://orcid.org/0000-0002-1617-6436}{\includegraphics[scale=0.06]{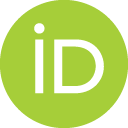}\hspace{1mm}
Hiya Gada}
\\
    Indian Institute of Technology Bombay\\
    Powai, Mumbai, 400 076, India \\
    \texttt{19d100007@iitb.ac.in} \\
\And	\href{https://orcid.org/0000-0003-4391-7014}{\includegraphics[scale=0.06]{orcid.png}\hspace{1mm}
Pieter van Goor}
\\
    Systems Theory and Robotics Group \\
	Australian National University \\
    ACT, 2601, Australia \\
    \texttt{Pieter.vanGoor@anu.edu.au} \\
	\And	\href{https://orcid.org/0000-0002-5746-7096}{\includegraphics[scale=0.06]{orcid.png}\hspace{1mm}
	Ravi Banavar}
\\
    Indian Institute of Technology Bombay\\
    Powai, Mumbai, 400 076, India \\
    \texttt{banavar@iitb.ac.in} \\
    \And	\href{https://orcid.org/0000-0002-7803-2868}{\includegraphics[scale=0.06]{orcid.png}\hspace{1mm}
    Robert Mahony}
\\
    Systems Theory and Robotics Group \\
	Australian National University \\
    ACT, 2601, Australia \\
	\texttt{Robert.Mahony@anu.edu.au} \\
}

\maketitle

\vspace{1cm}

\begin{abstract}
Observers for systems with Lie group symmetries are an active area of research that is seeing significant impact in a number of practical domains, including aerospace, robotics, and mechatronics.
This paper builds on the theory of the recently proposed Equivariant Filter (EqF), which is a general observer design for systems on homogeneous spaces that takes advantage of symmetries to yield significant performance advantages.
It is shown that the EqF error dynamics are invariant to transformation of the input signal and equivariant as a parametrised vector field.
The main theorem shows that two EqF's with different choices of local coordinates and origins and with equivalent noise modelling yield identical performance.
In other words, the EqF is intrinsic to the system equations and symmetry.
This is verified in a simulation of a 2D robot localisation problem, which also shows how the ability to choose an origin for the EqF can yield practical performance advantages by mitigating floating point precision errors.
\end{abstract} 

\section{Introduction}

Systems on Lie groups have been the subject of control research since the 1970s \citep{brockett1972system,jurdjevic1972control,brockett1973lie}.
In the 2000s, a number of authors in the nonlinear observer community studied systems on Lie groups for their application to attitude estimation, which was key to enabling the control of small unmanned aerial vehicles \citep{salcudean1991globally,thienel2003coupled,mahony2008nonlinear}.
Since then, the practical success of these observer designs has motivated the development of a general theory of observer design for systems on Lie groups.

Most early work on observer design for general systems on Lie groups sought to extend the results achieved for attitude estimation, and a number of approaches were taken by different authors.
\cite{bonnabel2006non} proposed one of the first frameworks for `invariant observers', and demonstrated their concept with an example application to velocity aided inertial navigation. 
\cite{bonnabel2007left} then developed a general theory of left-invariant extended Kalman filters (IEKFs) for systems on Lie groups, and demonstrated the effectiveness of the proposed architecture for the attitude estimation problem.
Following this, \cite{bonnabel2008symmetry} developed a theory of symmetry-preserving observers, and provided a constructive method for finding an invariant error and a symmetry-preserving correction term for a given system. 
In parallel work, \cite{lageman2009gradient} analysed invariant cost functions on Lie groups - as well as their construction from equivariant measurements - to propose a design methodology for observers for invariant systems on Lie groups with strong almost-global stability properties.
This was followed by \cite{mahony2013observers}, which introduced a distinction between a system with a Lie group symmetry and the corresponding lifted system on the group.
They proposed a design method for an observer for the lifted system on the Lie group, and showed that this respects the system geometry and can provide powerful stability guarantees.
\cite{bourmaud2013discrete} generalised a discrete extended Kalman filter (EKF) was to systems on Lie groups using a probabilistic perspective and a theory of concentrated Gaussian distributions.
Extending the work presented in \cite{mahony2013observers}, \cite{khosravian2015observers} proposed a new constructive observer design methodology for invariant systems on Lie groups with biased inputs.
\cite{saccon2015second} provided explicit formulas for the equations of a second-order-optimal minimum energy filter for systems on Lie groups by examining the Hamilton-Jacobi-Bellman equation associated with a loss function constructed from the input and output measurement disturbances of the system.

Recently, \cite{barrau2016invariant} and \cite{barrau2018invariant} carefully defined the IEKF and examined its convergence and stability properties for the special class of group-affine systems on Lie groups.
They defined an error function between the observer estimate and the system state which, in this special case of group-affine systems, was shown to evolve independently of the system and observer states.
Moreover, they showed that the evolution of this error function is exactly linear in the logarithmic coordinates of the Lie group.
\cite{van2021autonomous} also examined the observer design problem for group-affine systems, and developed a pre-observer that is synchronous with the system trajectories under the chosen definition of the error function.
\cite{joshi2019bundle} analysed the observer design problem for systems on general manifolds with a group action, and determined conditions under which the error dynamics are time-invariant and stable.
Finally, \cite{van2020equivariant} and \cite{mahony2021observer} developed the Equivariant Filter (EqF), which is a linearisation-based observer design for equivariant systems defined on manifolds with a transitive group action.  
They show how to define a global equivariant error, and how equivariance of the system output measurements may be exploited to improve the performance.

This paper examines symmetry properties of the EqF developed by \cite{van2020equivariant}.
Unlike other symmetry-based filters, the EqF is parametrised by a choice of origin in the system's state space manifold and local coordinates about that origin.
This freedom is particularly important for applications to systems on homogeneous spaces, where no natural choice of origin or local coordinates exists.
The EqF error dynamics are shown to be invariant to transformations of the velocity input and observer state, and equivariant as a parametrised vector field; that is, a symmetry applied to the vector field may be viewed as a symmetry on its parameters.
This implies that any change to the observer state may be viewed as a change to the input signal to the error dynamics, and characterises the effect of the observer state on the error dynamics entirely through a single group action.
The main result shows that any two EqF's for a given system, even with different choices for local coordinates and origins, yield identical performance when the modelling of system noise is equivalent.
Fundamentally, this means that the EqF design is intrinsic to the system equations and the symmetry; any choice of parametrisation leads to the same filter. 
This result is demonstrated in a simple simulation example involving localisation of a mobile robot in 2D space, where three different EqF's provide the same performance up to floating point precision errors.
These results demonstrate not only that the EqF is theoretically independent of the choice of local coordinates, but also that the freedom to change the local coordinates and origin of an EqF may be useful to practitioners, such as when developing an observer for localisation (and mapping) over a large operating environment or for applications requiring high precision. 




\section{Preliminaries}

For a clear introduction to smooth manifolds and Lie groups, the authors recommend \cite{lee2013smooth}.

Let $\calM, \calN$ be smooth manifolds.
The tangent space of $\calM$ at a point $\xi \in \calM$ is written $\tT_\xi \calM$, the tangent bundle is written $\tT \calM$, and the space of vector fields on $\calM$ is written $\gothX(\calM)$.
Given a function $h: \calM \to \calN$, the differential of $h$ with respect to $\zeta$ at $\xi \in \calM$ is denoted
\begin{align*}
    \at{\tD_\zeta}{\xi} h(\zeta) : \tT_\xi \calM &\to \tT_{h(\xi)} \calN, \\
    v &\mapsto \at{\tD_\zeta}{\xi} h(\zeta)[v].
\end{align*}
Alternatively, a shorter form denotes the differential as a map between the tangent bundles; that is,
\begin{align*}
    \tD h : \tT \calM &\to \tT \calN, \\
    v &\mapsto \tD h [v],
\end{align*}
where $v \in \tT_\xi \calM$ for some $\xi \in \calM$ that is left implicit.

Given two maps $h_1 : \calM \to \calN_1$, $h_2 : {\calN_1} \to \calN_2$, denote their composition by
\begin{align*}
    h_2 \circ h_1 : \calM &\to \calN_2, \\
    \xi &\mapsto h_2(h_1(\xi)).
\end{align*}
In the case of linear maps $H_1, H_2$, we may instead denote the composition using $H_2 \cdot H_1$, or simply $H_2 H_1$ to emphasise the relationship to matrix multiplication.
This is particularly useful in applications of the chain rule,
\begin{align*}
    \at{\tD_\zeta}{\xi} (h_2 \circ h_1) (\zeta) &= \at{\tD_\beta}{h_1(\xi)} h_2(\beta) \cdot \at{\tD_\zeta}{\xi} h_1(\zeta), \\
    \tD (h_2 \circ h_1) &= \tD h_2 \tD h_1.
\end{align*}

Denote a general Lie group $\grpG$, and its Lie algebra $\gothg$.
The Lie algebra may be identified with the tangent space of the group at the identity; i.e. $\gothg \simeq \tT_\id \grpG$.
Denote the Lie exponential map $\exp: \gothg \to \grpG$.
The left and right translation maps $L:\grpG \times \grpG \to \grpG$ and $R:\grpG \times \grpG \to \grpG$ are defined by 
\begin{align*}
    L_Z(X) &:= ZX, &
    R_Z(X) &:= XZ.
\end{align*}
The Adjoint map $\Ad: \grpG \times \gothg \to \gothg$ is defined by
\begin{align*}
    \Ad_X U := \at{\dds}{s=0} X \exp(s U) X^{-1}.
\end{align*}

A right\footnote{All group actions considered in this paper are right group actions.} group action of $\grpG$ on $\calM$ is a map $\phi : \grpG \times \calM \to \calM$ that satisfies
\begin{align*}
    \phi(\id, \xi) &= \xi, &
    \phi(Y, \phi(X, \xi)) &= \phi(X Y, \xi),
\end{align*}
for all $\xi \in \calM$ and all $X,Y \in \grpG$.
For any $\xi \in \calM$, the partial map $\phi_\xi : \grpG \to \calM$ is defined to be 
$$\phi_\xi(X) = \phi(X, \xi).$$
Likewise, for any $X \in \grpG$, the partial map $\phi_X : \calM \to \calM$ is defined to be 
$$\phi_X(\xi) = \phi(X, \xi).$$
A group action $\phi$ is said to be \emph{transitive} if, for all $\xi_1, \xi_2 \in \calM$, there exists $Z \in \grpG$ such that $\phi(Z, \xi_1) = \xi_2$.
The map $\Phi : \grpG \times \gothX(\calM) \to \gothX(\calM)$, defined by
\begin{align}
    \Phi_X(f)(\xi) := \tD_\zeta |_{\phi(X^{-1}, \xi)} \phi(X, \zeta) f(\phi(X^{-1}, \xi)), \label{eq:Phimap}
\end{align}
is a right action of $\grpG$ on $\gothX(\calM)$.

\subsection{Equivariant System and Lift}

A system on a manifold $\calM$ is defined by a smooth map $f : \vecL \to \gothX(\calM)$, which associates a vector field on $\calM$ to every input in the vector space $\vecL$.
Trajectories of the system are curves $\xi : [0, \infty) \to \calM$ satisfying 
$$\dot{\xi}(t) = f_{u(t)}(\xi(t)),$$
for all $t \geq 0$, where $u:[0,\infty) \to \vecL$ is a given input signal.
A measurement or output function is a smooth map 
$$h : \calM \to \calN \subset \R^n.$$

A symmetry of a system $f$ is a pair of group actions, $\phi : \grpG \times \calM \to \calM$ and $\psi : \grpG \times \vecL \to \vecL$, satisfying the equivariance property
\begin{align}
    \tD_\zeta |_\xi \phi_X (\zeta) \cdot f_u(\xi) &= f_{\psi_X(u)}(\phi_X(\xi)),
\end{align}
for all $X \in \grpG$, $\xi \in \calM$, and $u \in \vecL$.
In this case, we refer to $f$ as an \emph{equivariant system}, and the diagram

\[
\xymatrixcolsep{50pt}
\xymatrix{
\vecL
\ar[d]^{f}
\ar[r]^{\psi_X}
&
\vecL
\ar[d]^{f} 
\\
\gothX(\calM)
\ar[r]^{\Phi_X}
&
\gothX(\calM)
}
\]
commutes for all $X \in \grpG$, $\xi \in \calM$,
and $u \in \vecL$, where the map $\Phi_X$ is defined in \eqref{eq:Phimap}.


Given an equivariant system $f$ with symmetry $(\phi, \psi)$, there exists a Lie-algebraic function $\Lambda : \calM \times \vecL \to \gothg$ satisfying
\begin{align}
    \tD_{Z} |_\id \phi_\xi(Z) \Lambda(\xi, u) &= f_u(\xi), \\
    \Ad_{X^{-1}} \Lambda(\xi, u) &= \Lambda(\phi_X(\xi), \psi_X(u)),
\end{align}
for all $X \in \grpG$, $\xi \in \calM$, and $u \in \vecL$.
We refer to $\Lambda$ as an \emph{(equivariant) lift} of the system $f$ \citep{mahony2020equivariant}, and the following diagram commutes;
\[
\xymatrixcolsep{50pt}
\xymatrix{
\calM \times \vecL
\ar[d]^{\Lambda}
\ar[r]^{(\phi_X, \psi_X)}
&
\calM \times \vecL
\ar[d]^{\Lambda}
\\
\gothg
\ar[r]^{\Ad_{X^{-1}}}
&
\gothg
}
\]

\subsection{Equivariant Filter}

Consider an equivariant system $f$ with symmetry $(\phi, \psi)$, and let $\Lambda$ be an associated equivariant lift.
Consider a triple $(\hat{X}, \Sigma, \vartheta)$, where $\hat{X} \in \grpG$, $\Sigma \in \R^{m \times m}$, and $\vartheta : \calU \subset \calM \to \R^m$ is a local coordinate chart about a given origin $\mr{\xi} \in \calM$; i.e., $\vartheta(\mr{\xi}) = 0$.
The linearised state and output matrices associated with $(\hat{X}, \Sigma, \vartheta)$ are defined to be
\begin{align}
    \mr{A}_t
    &= \tD_{\zeta}|_{\mr{\xi}} \vartheta(\zeta) \cdot \tD_E |_\id \phi(E, \mr{\xi})
    \notag \\ &\phantom{==}
     \cdot \tD_\zeta |_{\mr{\xi}} \Lambda(\zeta, \psi(\hat{X}^{-1}, u)) \cdot \tD_x |_0 \vartheta^{-1} (x), \label{eq:Amat} \\
    C_t
    &= \tD_\xi |_{\phi(\hat{X}, \mr{\xi})} h(\xi) \cdot \tD_{\zeta}|_{\mr{\xi}} \phi(\hat{X}, \zeta) \cdot \tD_x |_0 \vartheta^{-1} (x),\label{eq:Cmat}
\end{align}
respectively \citep{van2020equivariant}.
The linearised state and output matrices, $A_t^\circ$ and $C_t$, given by the composition of maps in \eqref{eq:Amat} and \eqref{eq:Cmat} respectively are visualised better in the figure below that indicates the intermediate objects in the series of transformations.

\begin{tikzpicture}


\node[draw=none] (linearizedspace) at (0,0){$A_t^\circ: \quad \R^m$};

\node [draw=none]  (neighbourhood) at (3.5,0){${\tT_{\mr{\xi}} \calM}$};


\node [draw=none]  (linalg) at (7.5,0){$\gothg$};


\node [draw=none]  (vectorf) at (11,0){${\tT_{\mr{\xi}} \calM}$};


\node [draw=none]  (lins2) at (14,0){$\R^m$};

\draw[-stealth] (linearizedspace.east) -- (neighbourhood.west)
	node[midway,above]{$\tD_x |_0 \vartheta^{-1} (x)$};

\draw[-stealth] (neighbourhood.east) -- (linalg.west)
	node[midway,above]{$\tD_\zeta |_{\mr{\xi}} \Lambda(\zeta, \psi(\hat{X}^{-1}, u))$};

\draw[-stealth] (linalg.east) -- (vectorf.west)
	node[midway,above]{$\tD_E |_\id \phi(E, \mr{\xi})$};
	
\draw[-stealth] (vectorf.east) -- (lins2.west)
	node[midway,above]{$\tD_{\zeta}|_{\mr{\xi}} \vartheta(\zeta)$};

\end{tikzpicture}

\begin{tikzpicture}


\node[draw=none] (linearizedspace) at (0,0){${C_t} : \quad \R^m$};


\node [draw=none]  (vectorf) at (3.5,0){${\tT_{\mr{\xi}} \calM}$};


\node [draw=none]  (vectorf2) at (7,0){${\tT_{\phi(\hat{X}, \mr{\xi})} \calM}$};


\node [draw=none]  (lins2) at (10.5,0){$\R^n$};

\draw[-stealth] (linearizedspace.east) -- (vectorf.west)
	node[midway,above]{$\tD_x |_0 \vartheta^{-1} (x)$};
	
\draw[-stealth] (vectorf.east) -- (vectorf2.west)
	node[midway,above]{$\tD_{\zeta}|_{\mr{\xi}} \phi(\hat{X}, \zeta)$};
	
\draw[-stealth] (vectorf2.east) -- (lins2.west)
	node[midway,above]{$\tD_\xi |_{\phi(\hat{X}, \mr{\xi})} h(\xi)$};

\end{tikzpicture}

Let $\xi$ be a trajectory of the equivariant system $f$ with input signal $u = u(t) \in \vecL$ and output signal $y = h(\xi)$.
An Equivariant Filter (EqF) \citep{van2020equivariant} for this system is a triple $(\hat{X}, \Sigma, \vartheta)$ satisfying
\begin{align}
    \dot{\hat{X}} &= \hat{X} \Lambda(\phi(\hat{X}, \mr{\xi}), u) + \Delta \hat{X}, \label{eq:eqf_ode_observer} \\
    \Delta &= \tD_E |_\id \phi_{\mr{\xi}}(E)^{\dagger} \tD_x |_0 \vartheta^{-1} (x) \Sigma C_t^\top R (y - h(\phi(\hat{X}, \mr{\xi}))), \notag \\ 
    \dot{\Sigma} &= \mr{A}_t \Sigma + \Sigma {\mr{A}_t}^\top + Q - \Sigma C_t^\top R C_t \Sigma, \label{eq:eqf_ode_riccati}
\end{align}
where $\mr{A}_t, C_t$ are the state and output matrices associated with $(\hat{X}, \Sigma, \vartheta)$, respectively, $\tD_E |_\id \phi_{\mr{\xi}}(E)^{\dagger}$ is a right-inverse of $\tD_E |_\id \phi_{\mr{\xi}}(E)$, and $Q \in \Sym_+(m), R \in \Sym_+(n)$ are referred to as the state and output gain matrices, respectively.
The state estimate of an EqF $(\hat{X}, \Sigma, \vartheta)$ is defined to be $\hat{\xi} = \phi(\hat{X}, \mr{\xi})$.

\section{Equivariances of the EqF}

In this section, we consider an equivariant system $f$ with symmetry $(\phi, \psi)$, and a trajectory $\xi$ of that system with input signal $u \in \vecL$ and output signal $y=h(\xi) \in \calN \subset \R^n$.

\subsection{Equivariance of EqF Error Dynamics}

For an EqF $(\hat{X}, \Sigma, \vartheta)$, the global state error \citep{van2020equivariant} is defined to be $e = \phi(\hat{X}^{-1}, \xi)$ with dynamics 
\begin{align}
    \ddt e &= \chi(e; u, \hat{X}, \Delta, \mr{\xi}), \label{eq:error_dynamics} \\
    &= \tD_E |_\id \phi(E,e) \left[\Lambda(e, \psi_{\hat{X}^{-1}}(u)) - \Lambda(\mr{\xi}, \psi_{\hat{X}^{-1}}(u)) - \Delta\right], \notag
\end{align}
where $\mr{\xi} = \vartheta^{-1}(0)$ is the origin of the local coordinates.
The semicolon notation emphasises that $\chi$ may be thought of as a time-varying parametrised vector field, $\chi_{ (u, \hat{X},\Delta, \mr{\xi})} (e) \in \gothX(\calM)$ where the signals $(\mr{\xi}, u, \hat{X},\Delta)$ are treated as time-varying exogeneous parameters.
\begin{align*}
    \chi : \vecL \times \grpG \times \gothg \times \calM \to \gothX(\calM).
\end{align*}

\begin{proposition}
{The error dynamics parameterised by $(u, \hat{X}, \Delta, \mr{\xi})$ are invariant with
respect to a group action on $\vecL \times \grpG$, in the sense that}
\begin{align*}
    \chi(e; \psi(Z, u), {Z\hat{X}}, \Delta, \mr{\xi})
    = \chi(e; u, \hat{X}, \Delta, \mr{\xi}),
\end{align*}
for all $Z \in \grpG$.
That is, the following diagram commutes;
\[
\xymatrixcolsep{70pt}\xymatrix{
\vecL \times \grpG \times \gothg \times \calM
\ar[r]^{(\psi_Z, L_Z, \id_\gothg, \id_\calM)}
\ar[rd]^{\chi}
&
\vecL \times \grpG \times \gothg \times \calM
\ar[d]^{\chi}
\\
&
\gothX(\calM)
}
\]
\end{proposition}

where $\id_\gothg : \gothg \to \gothg$ and $\id_\calM : \calM \to \calM$ are the identity maps on $\gothg$ and $\calM$, respectively.


\begin{proof}
By direct computation,
\begin{align*}
    &\chi(e; u, Z \hat{X}, \Delta, \mr{\xi}) \\
    &= \tD \phi_e
    \left[\Lambda(e, \psi_{(Z\hat{X})^{-1}}(u))
    - \Lambda(\mr{\xi}, \psi_{(Z\hat{X})^{-1}}(u)) - \Delta\right], \\
    &= \tD \phi_e \left[\Lambda(e, \psi_{\hat{X}^{-1}}(\psi_{Z^{-1}}(u)))
    \right. \\ &\phantom{==} \left.
    - \Lambda(\mr{\xi}, \psi_{\hat{X}^{-1}}(\psi_{Z^{-1}}(u))) - \Delta\right], \\
    &= \chi(e; \psi(Z^{-1}, u), \hat{X}, \Delta, \mr{\xi}).
\end{align*}
\end{proof}

It follows that, in particular,
\begin{align}
    \chi(e; u, \hat{X}, \Delta, \mr{\xi})
    = \chi(e; \psi_{\hat{X}^{-1}}(u), \id, \Delta, \mr{\xi}).
\end{align}
In other words, $\hat{X}$ affects the dynamics of $e$ only through a transformation of the input signal $u$.

\begin{proposition} 
The error dynamics \eqref{eq:error_dynamics} are equivariant as a vector field, in the sense that

\begin{align*}
    {\Phi_Z \chi_{(u, \hat{X}, \Delta, \mr{\xi})}
    = \chi_{(\psi_Z(u), \hat{X}, \Ad_{Z^{-1}}\Delta, \phi_Z(\mr{\xi})}}
\end{align*}
for all $Z \in \grpG$.
That is, the following diagram commutes;
\[
\xymatrixcolsep{50pt}\xymatrix{
\vecL \times \grpG \times \gothg \times \calM
\ar[d]^{(\psi_Z, \id, \Ad_{Z^{-1}}, \phi_Z)}
\ar[r]^{\chi}
&
\gothX(M)
\ar[d]^{\Phi_Z}
\\
\vecL \times \grpG \times \gothg \times \calM
\ar[r]^{\chi}
&
\gothX(M)
}
\]
\end{proposition}

\begin{proof}
Observe that, for any $Z \in \grpG$, $\xi \in \calM$, and $\delta \in \gothg$,
\begin{align*}
    &\tD_\zeta |_\xi \phi(Z, \zeta) \cdot \tD_E |_\id \phi(E, \xi) [\delta] \\
    &= \at{\ddt}{t=0} \phi_\xi (\exp(t \delta)Z), \\
    &= \at{\ddt}{t=0} \phi(\exp(t \Ad_{Z^{-1}}\delta), \phi(Z, \xi)), \\
    &= \tD_E |_\id \phi(E, \phi(Z, e)) \cdot \Ad_{Z^{-1}} \delta,
\end{align*}
and thus $\tD_\zeta |_\xi \phi(Z, \zeta) \cdot \tD_E |_\id \phi(E, \xi) = \tD_E |_\id \phi(E, \phi(Z, \xi)) \cdot \Ad_{Z^{-1}}$.
Now, using the equivariance of the lift $\Lambda$,
\begin{align*}
    &\tD_e |_e \phi(Z, e) \cdot \chi(\phi_{Z^{-1}}(e); u, \hat{X}, \Delta, \mr{\xi}) \\
    &= \tD_e |_e \phi(Z, e) \cdot {\tD_E |_\id \phi(E,\phi_{Z^{-1}}(e))}
    \\ &\phantom{==}
    \left[\Lambda(\phi_{Z^{-1}}(e), \psi_{\hat{X}^{-1}}(u)) - \Lambda(\mr{\xi}, \psi_{\hat{X}^{-1}}(u)) - \Delta\right], \\
    &{=\tD_E |_\id \phi(E, \phi(Z, \phi_{Z^{-1}}(e)))} \cdot \Ad_{Z^{-1}}
    \\ &\phantom{==}
    \left[\Lambda(\phi_{Z^{-1}}(e), \psi_{\hat{X}^{-1}}(u)) - \Lambda(\mr{\xi}, \psi_{\hat{X}^{-1}}(u)) - \Delta\right], \\
    &= {\tD_E |_\id \phi(E, e)}
    \left[\Lambda(\phi(Z,\phi_{Z^{-1}}(e)), \psi_{Z} \circ \psi_{\hat{X}^{-1}}(u)) \right.
    \\ &\phantom{==}
     \left. - \Lambda(\phi(Z,\mr{\xi}), \psi_{Z} \circ \psi_{\hat{X}^{-1}}(u)) - \Ad_{Z^{-1}}\Delta\right], \\
    &{= \chi (e; u, Z^{-1}\hat{X}, \Ad_Z^{-1} \Delta, \phi_Z(\mr{\xi})),}\\
    &{= \chi (e; \psi_Z(u), \hat{X}, \Ad_Z^{-1} \Delta, \phi_Z(\mr{\xi})).}
\end{align*}
This completes the proof.
\end{proof}

\subsection{Eqvuiariance of the EqF Equations}

\begin{theorem} \label{theorem:origin_change}
Let $(\hat{X}, \Sigma, \vartheta)$ be an EqF with state and output gain matrices $Q,R$, respectively.
Suppose $\mr{\xi} \in \calM$ is the origin of $\vartheta$, i.e. $\vartheta(\mr{\xi}) = 0$.
Choose any $Z \in \grpG$, and choose a local coordinate chart $\bar{\vartheta}$ about $\ob{\mr{\xi}} := \phi(Z^{-1}, \mr{\xi})$.
If 
\begin{align}
    \bar{X} &:= Z \hat{X}, \\
    \bar{\Sigma} &:= M \Sigma M^\top, \\
    M &:= \tD_\zeta |_{\ob{\mr{\xi}}} \bar{\vartheta}(\zeta) \cdot \tD_\zeta |_{\mr{\xi}} \phi(Z^{-1}, \zeta) \cdot \tD_x|_0 \vartheta^{-1}(x),
\end{align}
then $(\bar{X}, \bar{\Sigma}, \bar{\vartheta})$ is an EqF with state and output gain matrices $M Q M^\top, R$, respectively.
\end{theorem}

\begin{proof}
For any $\varepsilon \in \R^m$, one has that
\begin{align*}
    &\tD_{\zeta}|_{\ob{\mr{\xi}}} \Lambda(\zeta, \psi(\bar{X}^{-1}, u)) \cdot \tD_x |_0 \bar{\vartheta}^{-1}(x) [\varepsilon] \\
    &= \at{\dds}{s=0} \Lambda(\bar{\vartheta}^{-1}(s\varepsilon), \psi(\bar{X}^{-1}, u)), \\
    &= \at{\dds}{s=0} \Ad_Z \Lambda(\phi_{Z}(\bar{\vartheta}^{-1}(s\varepsilon)), \psi(\hat{X}^{-1}, u)), \\
    &= \at{\dds}{s=0} \Ad_Z \Lambda(\vartheta^{-1} \circ \vartheta \circ \phi_{Z}(\bar{\vartheta}^{-1}(s\varepsilon)), \psi(\hat{X}^{-1}, u)), \\
    &= \Ad_Z \cdot \tD_{\zeta}|_{\phi_Z(\ob{\mr{\xi}})} \Lambda(\zeta, \psi(\hat{X}^{-1}, u)) \cdot
    \tD_x|_0 \vartheta^{-1} (x) 
    \\ &\phantom{==}
    \cdot \tD_\zeta |_{\mr{\xi}} \vartheta(\zeta) \cdot \tD_\zeta | _{\ob{\mr{\xi}}} \phi(Z, \zeta) \cdot \tD_x |_0 \bar{\vartheta}^{-1}(x)[\varepsilon]
    , \\
    &= \Ad_Z \cdot \tD_{\zeta}|_{\phi_Z(\ob{\mr{\xi}})} \Lambda(\zeta, \psi(\hat{X}^{-1}, u)) \cdot
    \tD_x|_0 \vartheta^{-1} (x) \cdot M^{-1} [\varepsilon].
\end{align*}
Similarly, for any $\delta \in \gothg$,
\begin{align*}
    &\tD_\zeta |_{\ob{\mr{\xi}}} \bar{\vartheta}(\zeta) \cdot \tD_E |_\id \phi(E, \ob{\mr{\xi}}) \cdot \Ad_Z [\delta] \\
    &= \at{\dds}{s=0} \bar{\vartheta} \circ \phi(e^{s\Ad_Z \delta}, \ob{\mr{\xi}}), \\
    &= \at{\dds}{s=0} \bar{\vartheta} \circ \phi_{Z^{-1}} \circ \phi(e^{s \delta}, \phi(Z , \ob{\mr{\xi}})), \\
    &= \at{\dds}{s=0} \bar{\vartheta} \circ \phi_{Z^{-1}} \circ \vartheta^{-1} \circ \vartheta \circ \phi(e^{s \delta}, \mr{\xi}), \\
    &= \tD_\zeta |_{\ob{\mr{\xi}}} \bar{\vartheta}(\zeta)
    \cdot \tD_\zeta |_{\mr{\xi}} \phi(Z^{-1}, \zeta)
    \cdot \tD_x|_0 \vartheta^{-1}(x)
    \\ &\phantom{==}
    \cdot \tD_\zeta |_{\mr{\xi}} \vartheta(\zeta)
    \cdot \tD_E |_\id \phi(E, \mr{\xi})[\delta], \\
    &= M \cdot \tD_\zeta |_{\mr{\xi}} \vartheta(\zeta) \tD_E |_\id \phi(E, \mr{\xi})[\delta].
\end{align*}
Now, let $\bar{A}_t$ denote the state matrix associated with $(\bar{X}, \bar{\Sigma}, \bar{\vartheta})$.
Then,
\begin{align*}
    &\bar{A}_t
    = \tD_\zeta |_{\ob{\mr{\xi}}} \bar{\vartheta}(\zeta ) \cdot \tD_E |_\id \phi(E, \ob{\mr{\xi}})
    \\ &\phantom{==}
    \cdot \tD_{\zeta}|_{\ob{\mr{\xi}}} \Lambda(\zeta, \psi(\bar{X}^{-1}, u)) \cdot \tD_x |_0 \bar{\vartheta}^{-1}(x), \\
    &= \tD_\zeta |_{\ob{\mr{\xi}}} \bar{\vartheta}(\zeta ) \cdot \tD_E |_\id \phi(E, \ob{\mr{\xi}})
    \\ &\phantom{==}
    \cdot \Ad_Z \cdot \tD_{\zeta}|_{\phi_Z(\ob{\mr{\xi}})} \Lambda(\zeta, \psi(\hat{X}^{-1}, u)) \cdot
    \tD_x|_0 \vartheta^{-1} (x) \cdot M^{-1}, \\
    &=  M \cdot \tD_\zeta |_{\mr{\xi}} \vartheta(\zeta) \tD_E |_\id \phi(E, \mr{\xi})
    \\ &\phantom{==}
    \cdot \tD_{\zeta}|_{\phi_Z(\ob{\mr{\xi}})} \Lambda(\zeta, \psi(\hat{X}^{-1}, u)) \cdot
    \tD_x|_0 \vartheta^{-1} (x) \cdot M^{-1}, \\
    &=  M \mr{A}_t M^{-1},
\end{align*}
where $\mr{A}_t$ is the state matrix associated with $(\hat{X}, \Sigma, \vartheta)$.

Let $\bar{C}_t$ denote the output matrix associated with $(\bar{X}, \bar{\Sigma}, \bar{\vartheta})$.
Then for any $\varepsilon \in \R^m$,
\begin{align*}
    \bar{C}_t
    &= \tD_\xi |_{\phi(\bar{X}, \ob{\mr{\xi}})} h(\xi) \cdot \tD_\zeta |_{\ob{\mr{\xi}}} \phi(\bar{X}, \zeta) \cdot \tD_x |_0 \bar{\vartheta}^{-1} (x)[\varepsilon], \\
    &= \at{\dds}{s=0} h \circ \phi(\bar{X}, \bar{\vartheta}^{-1} (s \varepsilon)), \\
    &= \at{\dds}{s=0} h \circ \phi(\hat{X}, \phi_Z \circ \bar{\vartheta}^{-1} (s \varepsilon)), \\
    &= \at{\dds}{s=0} h \circ \phi(\hat{X}, \vartheta^{-1} \circ \vartheta \circ \phi_Z \circ \bar{\vartheta}^{-1} (s \varepsilon)), \\
    &= \tD_\xi |_{\phi(\hat{X}, \mr{\xi})} h(\xi)
    \cdot \tD_\zeta |_{\mr{\xi}} \phi(\hat{X}, \zeta)
    \cdot \tD_x |_0 \vartheta^{-1} (x)
    \\ &\phantom{==}
    \cdot \tD_\zeta |_{\mr{\xi}} \vartheta(\zeta)
    \cdot \tD_\zeta | _{\ob{\mr{\xi}}} \phi(Z, \zeta)
    \cdot \tD_x |_0 \bar{\vartheta}^{-1}(x)[\varepsilon], \\
    &= C_t M^{-1} \varepsilon,
\end{align*}
where $C_t$ is the output matrix associated with $(\hat{X}, \Sigma, \vartheta)$.

From here the derivative of $\bar{\Sigma}$ may be computed by
\begin{align*}
    \dot{\bar{\Sigma}}
    &= M \dot{\Sigma} M^\top, \\
    &= M (\mr{A}_t \Sigma + \Sigma {\mr{A}_t}^\top + Q - \Sigma C_t^\top R C_t \Sigma) M^\top, \\
    &= M \mr{A}_t \Sigma M^\top + M \Sigma {\mr{A}_t}^\top M^\top 
    \\ &\phantom{==}
    + M Q M^\top - M \Sigma C_t^\top R C_t \Sigma M^\top, \\
    &= M \mr{A}_t M^{-1} M \Sigma M^\top + M \Sigma M^\top M^{-\top} {\mr{A}_t}^\top M^\top 
    \\ &\phantom{==}
    + M Q M^\top - M \Sigma M^\top M^{-\top} C_t^\top R C_t M^{-1} M \Sigma M^\top, \\
    &= \bar{A}_t \bar{\Sigma} + \bar{\Sigma} \bar{A}_t^\top 
    + M Q M^\top - \bar{\Sigma} \bar{C}_t R \bar{C}_t \bar{\Sigma}.
\end{align*}
Thus $\bar{\Sigma}$ satisfies the EqF Riccati equation \eqref{eq:eqf_ode_riccati} for the state matrix and output matrix associated with $(\bar{X}, \bar{\Sigma}, \bar{\vartheta})$, and with state and output gain matrices $M Q M^\top$ and $R$, respectively.

Observe that, for any $v \in \tT_{\mr{\xi}} \calM$,
\begin{align*}
    &\tD_E |_\id \phi_{\ob{\mr{\xi}}}(E) \cdot \Ad_Z 
    \cdot \tD_E |_\id \phi_{\mr{\xi}}(E)^\dagger [v] \\
    &= \at{\dds}{s=0} \phi(Z \exp( s \tD_E |_\id \phi_{\mr{\xi}}(E)^\dagger [v]) Z^{-1}, \ob{\mr{\xi}}), \\
    &= \at{\dds}{s=0} \phi_Z \circ \phi(\exp( s \tD_E |_\id \phi_{\mr{\xi}}(E)^\dagger [v]), \phi(Z, \ob{\mr{\xi}})), \\
    &= \tD_\zeta |_{\mr{\xi}} \phi(Z, \zeta)
    \cdot \tD_E |_\id \phi(E, \mr{\xi})
    \cdot \tD \tD_E |_\id \phi_{\mr{\xi}}(E)^\dagger [v], \\
    &= \tD_\zeta |_{\mr{\xi}} \phi(Z, \zeta) [v].
\end{align*}
Hence $\tD_E |_\id \phi(E, \ob{\mr{\xi}})^\dagger := \Ad_Z \tD_E |_\id \phi_{\mr{\xi}}(E)^\dagger \tD_\zeta |_{\ob{\mr{\xi}}} \phi(Z^{-1}, \zeta)$ is a right-inverse of $\tD_\zeta |_{\ob{\mr{\xi}}} \phi_{\ob{\mr{\xi}}}(\id)$.
Now, computing the derivative of $\bar{X}$ yields
\begin{align*}
    \dot{\bar{X}}
    &= Z \dot{\hat{X}}, \\
    &= Z (\hat{X} \Lambda(\phi(\hat{X}, \mr{\xi}), u) + \Delta \hat{X}), \\
    &= Z \hat{X} \Lambda(\phi(\hat{X}, \mr{\xi}), u) + Z \Delta Z^{-1} Z \hat{X}, \\
    &= \bar{X} \Lambda(\phi(\hat{X}, \phi(Z, \ob{\mr{\xi}})), u) + \Ad_Z \Delta \bar{X}, \\
    &= \bar{X} \Lambda(\phi(\bar{X}, \ob{\mr{\xi}}), u) + \bar{\Delta} \bar{X},
\end{align*}
where,
\begin{align*}
    \bar{\Delta} &= \Ad_Z \Delta, \\
    &= \Ad_Z \tD_E |_\id \phi(E, \mr{\xi})^{\dagger} \tD_x |_0 \vartheta^{-1}(x)
    \\ &\phantom{==}
    \cdot \Sigma C_t^\top R (y - h(\phi(\hat{X}, \mr{\xi}))), \\
    &= \Ad_Z \tD_E |_\id \phi(E, \mr{\xi})^{\dagger} \tD_x |_0 \vartheta^{-1}(x) M^{-1}
    \\ &\phantom{==}
    \cdot \bar{\Sigma} \bar{C}_t^\top R (y - h(\phi(\hat{X}, \phi(Z, \ob{\mr{\xi}})))), \\
    &= \tD_E |_\id \phi(E, \ob{\mr{\xi}})^\dagger \tD_x |_0 \bar{\vartheta}^{-1}(x) \bar{\Sigma} \bar{C}_t^\top R (y - h(\phi(\bar{X}, \ob{\mr{\xi}}))).
\end{align*}
This means that $\bar{X}$ satisfies exactly the EqF observer equation \eqref{eq:eqf_ode_observer} associated with $(\bar{X}, \bar{\Sigma}, \bar{\vartheta})$.
Therefore, $(\bar{X}, \bar{\Sigma}, \bar{\vartheta})$ is an EqF with state and output gain matrices $M Q M^\top$ and $R$, respectively, and this completes the proof.
\end{proof}

\begin{corollary}\label{cor:equivalent_eqfs}
Let $(\hat{X}, \Sigma, \vartheta)$ be an EqF with origin $\mr{\xi}$ and state and output gain matrices $Q,R$, respectively.
Likewise, let $(\bar{X}, \bar{\Sigma}, \bar{\vartheta})$ be an EqF with origin $\ob{\mr{\xi}}$ and state and output gain matrices $\bar{Q},\bar{R}$, respectively.
Suppose that
\begin{gather*}
    \begin{align}
        \phi(\bar{X}(0), \ob{\mr{\xi}}) &= \phi(\hat{X}(0), \mr{\xi}), &
        \bar{\Sigma}(0) &= M \Sigma(0) M^\top, \\
        \bar{Q} &= M Q M^\top, &
        \bar{R} &= R,
    \end{align} \\
    \tD_E |_\id \phi(E, \ob{\mr{\xi}})^\dagger := \Ad_Z \tD_E |_\id \phi_{\mr{\xi}}(E)^\dagger \tD_\zeta |_{\ob{\mr{\xi}}} \phi(Z^{-1}, \zeta),
\end{gather*}
where
\begin{align*}
    Z &:= \bar{X}(0) \hat{X}(0)^{-1}, \\
    M &:= \tD_\zeta |_{\ob{\mr{\xi}}} \bar{\vartheta}(\zeta) \cdot \tD_\zeta |_{\mr{\xi}} \phi(Z^{-1}, \zeta) \cdot \tD_x|_0 \vartheta^{-1}(x).
\end{align*}
Then $\bar{X}(t) = Z \hat{X}(t)$ and $\bar{\Sigma}(t) = M \Sigma(t) M^\top$ for all time $t \geq 0$.
\end{corollary}

\begin{proof}
From Theorem \ref{theorem:origin_change}, the triple $(Z \hat{X}, M \Sigma M^\top, \bar{\vartheta})$ is an EqF with origin $\ob{\mr{\xi}}$ and state and output gains matrices $M Q M^\top, R$, respectively.
Moreover, one has that
\begin{align*}
    Z \hat{X}(0) &= \bar{X}(0) \hat{X}(0)^{-1} \hat{X}(0) = \bar{X}(0).
\end{align*}
Thus, the initial conditions of the EqF $(Z \hat{X}, M \Sigma M^\top, \bar{\vartheta})$ and those of $(\bar{X}, \bar{\Sigma}, \bar{\vartheta})$ are equal.
Since these EqF's also share local coordinates and gain matrices, it follows that $(Z \hat{X}, M \Sigma M^\top, \bar{\vartheta})$ satisfies the ODE defining $(\bar{X}, \bar{\Sigma}, \bar{\vartheta})$.
Therefore, by uniqueness of the solution of the initial value problem, 
\begin{align*}
    \bar{X}(t) &= Z \hat{X}(t), &
    \bar{\Sigma}(t) &= M \Sigma(t) M^\top,
\end{align*}
for all time $t \geq 0$, as required.
\end{proof}

\section{Example: 2D localisation}


We verify the general theory presented by showing a simple example of 2D robot localisation from landmark measurements.
Consider a robot moving in the 2D plane, equipped with a sensor (e.g. a depth camera or LIDAR) able to recognise some known landmarks.
For simplicity, assume that all landmarks are visible at all times, and that the robot's rigid body velocity can be measured.

\subsection{System Definition}

The Special Euclidean group $\SE(2)$ is a matrix Lie group,
\begin{align*}
    \SE(2) := \left\{ \begin{pmatrix} R & x \\ 0 & 1 \end{pmatrix} \mid R \in \SO(2), x \in \R^2  \right\},
\end{align*}
where $R \in \SO(2)$ is a 2D rotation matrix of the form
\begin{align*}
    R(\theta) = \begin{pmatrix} \cos(\theta) & -\sin(\theta) \\ \sin(\theta) & \cos(\theta) \end{pmatrix},
\end{align*}
for some angle $\theta \in [0, 2\pi)$.
The Lie algebra of $\SE(2)$ is denoted $\se(2)$.
Elements of $\SE(2)$ may be written $P = (R_P, x_P)$, where $R_P = R(\theta_P)$ is the rotation component of $P$ and $x_P$ the translation component of $P$.

The system dynamics $f: \se(2) \to \gothX(\SE(2))$
and the system measurement $h: \SE(2) \to \R^{2n}$ are defined to be
\begin{align*}
    f_U(P) &:= P U, \\
    h(P) &:= (R_P^\top(p_1 - x_P), ..., R_P^\top(p_n - x_P)),
\end{align*}
where the robot measures the positions of $n$ known landmarks $p_1,..., p_n \in \R^2$.

{In this particular example, we observe the following equivalence with our previous notation, 
\begin{align*}
    \calM &= \SE(2), &
    \vecL &= \se(2), &
    \grpG &= \SE(2).
\end{align*}} 
Define the group actions $\phi: \SE(2) \times \SE(2) \to \SE(2)$ and $\psi: \SE(2) \times \se(2) \to \se(2)$ by
\begin{align*}
    \phi(X, P) &:= PX, &
    \psi(X, U) &:= X^{-1} U X.
\end{align*}
It is trivial to verify that both of these maps are group actions, that $\phi$ is transitive, and that $f$ is equivariant with respect to $(\phi, \psi)$.

Define $\Lambda : \SE(2) \times \se(2) \to \se(2)$,
\begin{align*}
    \Lambda(P, U) = U.
\end{align*}
Then for any $P \in \SE(2)$, $U \in \se(2)$ and $X \in \SE(2)$,
\begin{align*}
    \tD_Z |_\id \phi_P(Z) \Lambda(P, U)
    = \tD_Z |_\id \phi_P(Z) [U]
    = P U,
\end{align*}
and,
\begin{align*}
    \Ad_{X^{-1}} \Lambda(P, U)
    = X^{-1} U X
    = \Lambda(\phi_X(P), \psi_X(U)).
\end{align*}
Thus $\Lambda$ is an equivariant lift for $f$.

\subsection{Simulation Results}

The initial orientation and position of the robot were chosen to be $P(0) = (\theta_P(0), x_P(0)) = (0, (0.7 \; 0.5)^\top)$, and the velocity was given by $\Omega_U = {0.4}$~rad/s, $V_U = (0.5 \; 0)^\top$~m/s.
{The robot moves in a circular trajectory in the anticlockwise direction and covers about $8$~rad.}
Five landmarks were considered, with positions randomly generated from $p_i \sim \calN(0, I_2)$.
The system was integrated for $20$~s, using Euler integration with an integration step of $\td t = 0.1$~s.

The triples $(\hat{X}_i, \Sigma_i, \vartheta_i)$ were defined to satisfy the EqF equations (\ref{eq:eqf_ode_observer},\ref{eq:eqf_ode_riccati}) with local coordinates given by
\begin{align*}
    \mr{\xi}_1 = \vartheta_1^{-1}(0) &= (I_2, 10^3[1 \; 1]^\top), \\
    \mr{\xi}_2 = \vartheta_2^{-1}(0) &= (I_2, 10^4[1 \; 1]^\top), \\
    \mr{\xi}_3 = \vartheta_3^{-1}(0) &= (I_2, 10^5[1 \; 1]^\top), \\
    \vartheta_i(P) &= \begin{pmatrix} \theta_P - \theta_{\mr{\xi}_i} & x_P - x_{\mr{\xi}_i} & y_P - y_{\mr{\xi}_i}  \end{pmatrix}^\top.
\end{align*}
The state gain $Q_i$, output gain $R_i$, and initial Riccati matrices $\Sigma_i(0)$ of the EqF's were chosen to satisfy the assumptions of Corollary \ref{cor:equivalent_eqfs}.
 
\begin{figure}[ht]
    \centering
    \includegraphics[width=\figsize\linewidth]{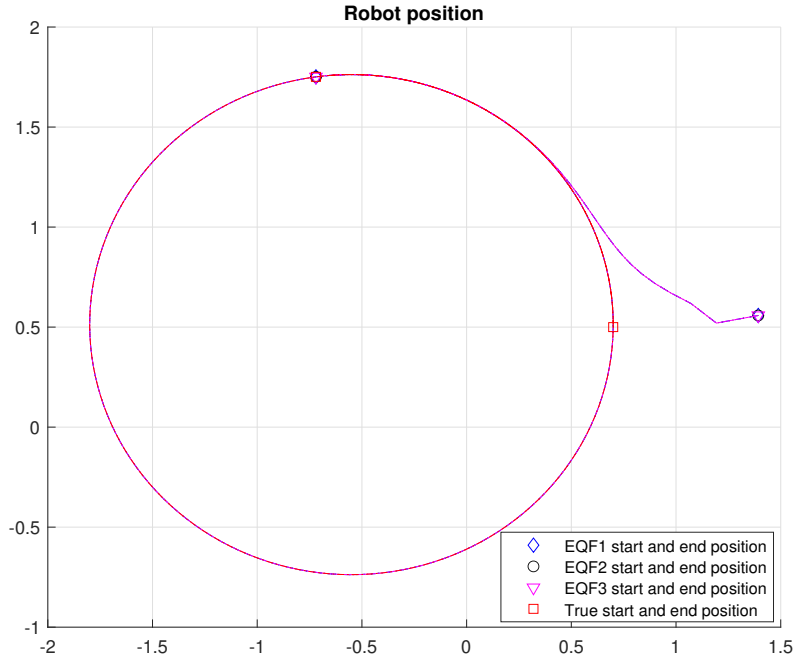}
    \caption{The true and estimated robot position trajectory of three EqF's with different local coordinates and origins.
    The overlapping trajectories verify that, indeed, the EqF's yield identical estimation performance.}
    \label{fig:trajectories}
\end{figure}

\begin{figure}[ht]
    \centering
    \includegraphics[width=\figsize\linewidth]{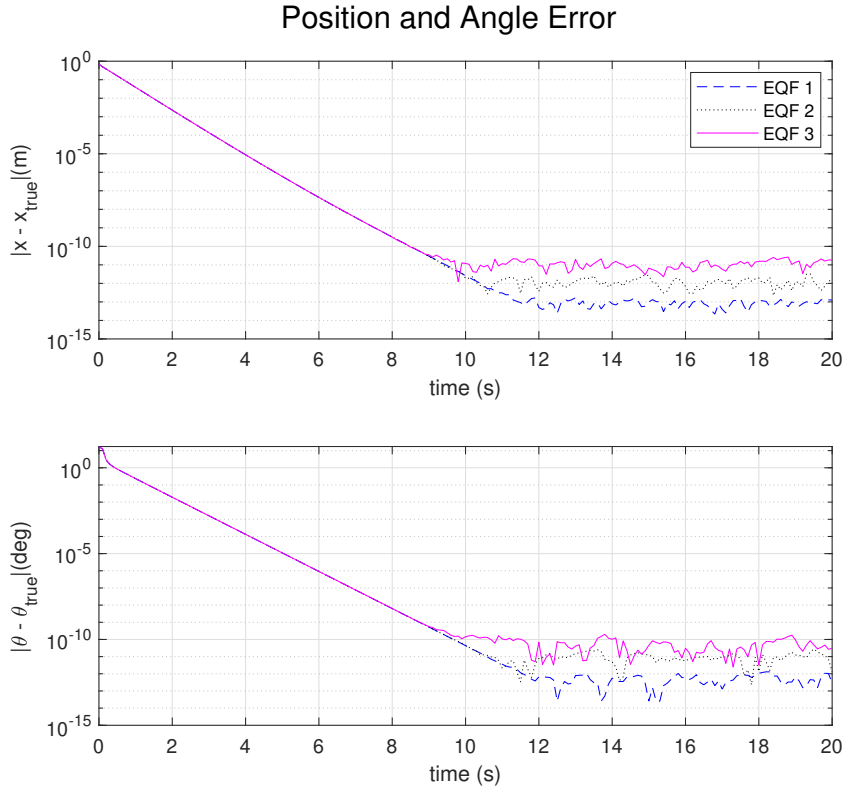}
    \caption{The absolute errors between the true and estimated position and angles of three EqF's with different local coordinates and origins.
    The EqF's with more distant origins are affected by numerical precision errors sooner, demonstrating the value of the freedom to choose an origin.}
    \label{fig:errors}
\end{figure}

Figure \ref{fig:trajectories} shows the true and estimated positions of each EqF over time, and Figure \ref{fig:errors} show the absolute differences between the true and estimated positions and angles of the robot.
The overlapping trajectories in Figure \ref{fig:trajectories} show that, generally, the three EqF's offer identical performance in estimation of the robot position.
Figure \ref{fig:errors}, however, shows that the effects of numerical precision occur sooner for the EqF's with more extreme choices of origin $\mr{\xi}_i$.
These results verify the theory that each EqF should yield the same state estimate provided equivalent tuning.
Additionally, they demonstrates the value of the freedom to choose an origin to a practitioner, as a potential tool to overcome floating point precision issues, such as in systems operating in large environments or those with high precision.

\section{Conclusion}

This paper examines the symmetry properties of the recently proposed Equivariant Filter (EqF) \citep{van2020equivariant}.
The EqF error dynamics are shown to be invariant to transformations of the input signal and observer state, and equivalently, the effect on the error dynamics of a disturbance to the observer state is shown to be entirely characterised by a group action on the input space.
The EqF error dynamics are additionally shown to be equivariant as a vector field to a change of the underlying parameters.
The main result shows that two EqF's for a given system with equivalent noise modelling yield identical estimates; that is, the filter is intrinsic (independent of local coordinates and origin choices).
An example simulation verifies this, and moreover shows how choosing a different origin for an EqF may yield practical performance advantages in the presence of floating point precision errors.


\bibliographystyle{plainnat}    
\bibliography{EqF}
\end{document}